\documentclass[%
twocolumn,secnumarabic,amssymb, amsmath, tightenlines,nobibnotes, aps, prl,longbibliography]{revtex4-1}

\usepackage[pdftex,colorlinks=true,pdfstartview=Fit]{hyperref}		

\usepackage[normalem]{ulem}		

\usepackage{bm}		
\usepackage{amssymb,amsmath}
\usepackage{graphicx}

\usepackage{epstopdf}
\usepackage{pdfpages}
\usepackage{etoolbox} 

\makeatletter
\patchcmd{\@outputpage@head}{\@ifx{\LS@rot\@undefined}{}{\LS@rot}}{}{}{}
\makeatother


\begin{document}

\title{Direct measurement of nonequilibrium system entropy is consistent with Gibbs-Shannon form}


\author{Mom\v cilo Gavrilov$^{1,\dag}$, Rapha\"el Ch\'etrite$^{1,2,3}$, and John Bechhoefer$^{1}$}
\email[email: ]{johnb@sfu.ca}

\affiliation{$^{1}$ Department of Physics, Simon Fraser University, Burnaby, British Columbia, V5A 1S6, Canada\\
	$^{2}$ Pacific Institute for the Mathematical Sciences,  UMI 3069, Vancouver, British Columbia, Canada \\
	$^{3}$ Universit\'e C\^ote d'Azur, CNRS, LJAD, Parc Valrose, 06108 NICE Cedex 02, 
	France \\
	$\dag$Present address:  Department of Biophysics and Biophysical Chemistry, Johns Hopkins University, 725 N. Wolfe Street, Baltimore, MD 21205-2185, USA}

\begin{abstract}
	Stochastic thermodynamics extends classical thermodynamics to small systems in contact with one or more heat baths.  It can account for the effects of thermal fluctuations and describe systems far from thermodynamic equilibrium.  A basic assumption is that the expression for Shannon entropy is the appropriate description for the entropy of a nonequilibrium system in such a setting.  Here, for the first time, we measure experimentally this function.  Our system is a micron-scale colloidal particle in water, in a virtual double-well potential created by a feedback trap.  We measure the work to erase a fraction of a bit of information and show that it is bounded by the Shannon entropy for a two-state system.  Further, by measuring directly the reversibility of slow protocols, we can distinguish unambiguously between protocols that can and cannot reach the expected thermodynamic bounds.
\end{abstract}

\maketitle

\section{Introduction}
Beginning with the foundational work of Clausius, Maxwell, and Boltzmann in the 19th~c., the concept of entropy has played a key role in thermodynamics.  Yet, despite its importance, entropy is an elusive concept \cite{grad61, penrose70, wehrl78, mackey89, wehrl91, balian03, jaynes83, bricmont96}, with no unique definition; rather, the appropriate definition of entropy depends on the scale, relevant thermodynamic variables, and nature of the system, with ongoing debate existing over the proper definition even for equilibrium cases \cite{dunkel14}.  Moreover, entropy has not been directly measured but is rather inferred from other quantities, such as the integral of the specific heat divided by temperature.  Here, by measuring the work required to erase a fraction of a bit of information, we isolate directly the change in entropy in an open  nonequilibrium system, showing that it is compatible with the functional form proposed by Gibbs and Shannon, giving it a physical meaning in this context.   Knowing the relevant form of entropy is crucial for efforts to extend thermodynamics to systems out of equilibrium.


For a continuous classical system whose state in phase space $x$ is distributed as the probability density function $\rho(x)$, the Gibbs-Shannon entropy is \cite{gibbs1902,shannon48} 
\begin{equation}
	S = -k_{\rm B} \int {\rm d}x \, \rho(x) \, \ln \rho(x) \,,
\label{eq:GibbsShannon}
\end{equation}
where $k_{\rm B}$ is Boltzmann's constant.  For quantum systems, von Neumann introduced, in 1927, the corresponding expression in terms of the density matrix \cite{vonNeumann27}.  Historically, the system in Eq.~\ref{eq:GibbsShannon} has typically been assumed to be in thermal equilibrium.  

The physical relevance of Eq.~\ref{eq:GibbsShannon} for a nonequilibrium distribution $\rho(x)$ has often been questioned (e.g., \cite{skagerstam75,lebowitz93,goldstein04,hemmo12,dahlsten11}).  One concern is that $S$ is constant on an isolated Hamiltonian system and can change only when evaluated on subsystems, such as those picked out by coarse graining. With many ways to choose subsystems or to coarse grain, is the associated notion of irreversibility intrinsic to the description of the system?
 
In another approach to entropy, advanced in the context of communication and information theory, Shannon \cite{shannon48,cover06} proved that, up to a  multiplicative constant, $S$ is the only possible function satisfying three intuitive axioms.  Alternatively, one can start from an axiomatic framework for thermodynamics \cite{lieb13,weilenmann16}.  The importance of using the appropriate form of entropy is highlighted in the recently developed field of \textit{stochastic thermodynamics} \cite{jarzynski97,crooks98,lebowitz99a,maes99,maes09,jarzynski00,seifert05,chetrite08,sekimoto10,seifert12}, where a central, underlying  hypothesis is that Eq.~\ref{eq:GibbsShannon} applies to densities defined for nonequilibrium mesoscopic systems coupled to one or more heat baths \cite{seifert05}.  We emphasize that this extension of the equilibrium Gibbs-Shannon entropy to nonequilibrium systems remains controversial within part of the statistical physics community, mainly for the reason that it is constant for Hamiltonian systems.

In this letter, we offer an experimental approach:  in a nonequilibrium system, we  measure directly the change in the entropy of the system and show that it is compatible with the postulated Gibbs-Shannon form, Eq.~\ref{eq:GibbsShannon}.

Our system is a micron-scale silica bead in water at temperature $T$ that serves as a reservoir, or heat bath.  We use a  feedback trap  \cite{cohen05d}  to create a virtual symmetric double-well potential $U(x,t)$ that models a one-bit memory.  The particle motion in this trap obeys nearly ideal Langevin (Brownian) overdamped dynamics \cite{gardiner09,risken89,vanKampen07}, \begin{equation}
	\dot{x}(t) = -\frac{1}{\gamma} \left. \frac{\partial U(x,t)}{\partial x} \right|_{x(t)} + \sqrt{\frac{2k_{\rm B}T}{\gamma}} \, \nu(t) \,,
\label{eq:langevin}
\end{equation}
where $\nu(t)$ denotes white-noise forcing, Gaussian with unit variance, and $\gamma$ denotes the damping.

We show that erasing a fraction of this bit requires, from a generalization of the Landauer principle for a two-state system \cite{landauer61}, a minimal average work whose value is set by the Gibbs-Shannon system entropy given in Eq.~\ref{eq:GibbsShannon}.  Our main goal, however, is not to further explore Landauer's principle but rather to \textit{use} it to test whether the Shannon entropy has a physical meaning in the nonequilibrium contexts probed by our experiments.

Appropriate experimental protocols require complex, precise control of the shape of the potential, $U(x,t)$.  Such control---involving barrier height, tilt, and local coordinate stretching to produce asymmetry between macrostates---is easy to achieve using feedback traps, where the form of a ``virtual potential'' is defined in software by applying the force that would be applied by a physical potential (see Methods).  By contrast, it is very difficult to achieve using an ordinary, physical potential.  Combining those operations, we construct thermodynamically reversible protocols that can reach theoretical bounds for required work in the slow limit.

\section*{Theory}

\subsection{Second law of thermodynamics in terms of work}

The second law of thermodynamics asserts that during a time interval $[0,\tau]$, the entropy production $S_{\rm tot} \ge 0$ \cite{penrose70,kondepudi14,degroot62,maes03,maes09}.  This entropy production is that of the total system, including the surrounding medium (heat bath), and decomposes into two terms:
\begin{equation}
	S_{\rm tot} = S_{\rm m} + \Delta S \,,
\label{eq:secondLaw1}
\end{equation}
where $S_{\rm m}$ is the entropy exchanged with the surrounding medium and where $\Delta S = S_\tau - S_0$ is the difference in system entropy over the time interval.  At this point, $S_0$ and $S_\tau$ are not necessarily given by the Shannon entropy.  Using the Clausius principle (1850) for the equilibrium bath \cite{clausius1850}, we can write $S_{\rm m} = Q / T$, where $Q$ is the heat exchanged with the medium, defined to be positive if the transfer is from the system to the medium.  Mathematically, the equilibrium character of the bath is reflected by the fact that the amplitude $2k_{\rm B}T$ in front of the noise term in the Langevin equation, Eq.~\ref{eq:langevin}, is constant and well defined during the entire protocol.  Physically, this hypothesis means that the time scales of the particle are much slower than those of the bath. 

The second law then becomes $Q \ge - T \, \Delta S$.  To reformulate the second law in terms of work, we use the first law, 
\begin{equation}
	W = \Delta E + Q \,,
\label{eq:firstLawStochastic}
\end{equation}
where $W$ is the average work done on the system to carry out the protocol over time $\tau$.  In the context of stochastic thermodynamics for overdamped dynamics, Eq.~\ref{eq:langevin}---small systems in contact with a large bath---work is calculated using the average of the Sekimoto formula (Eq.~\ref{eq:sekimotoWork} in methods) \cite{sekimoto97,jarzynski97,jarzynski00,sekimoto10}.  Then, using the nonequilibrium free energy $F_{\rm neq} = E - T S$, the expression for heat $Q$ given above, and Eq.~\ref{eq:firstLawStochastic}, we have 
\cite{esposito11,parrondo15}
\begin{equation}
	W \ge \Delta F_{\rm neq} \,.
\label{eq:secondLaw3}
\end{equation}
Note that the average energy $E$ at time $t$ is determined from the potential $U(x,t)$ and the instantaneous density of the process $\rho(x,t)$ by
\begin{equation}
	E(t) = \int_{-\infty}^\infty {\rm d}x \, \rho(x,t) \, U(x,t) \,.
\label{eq:Eavg}
\end{equation}
The nonequilibrium free energy $F_{\rm neq}$ reduces to the conventional equilibrium free energy, defined using the partition function, when the average energy $E$ and entropy $S$ are evaluated from equilibrium distributions.

\subsection{Coarse graining from a continuous to a discrete system}
\label{sec:coarse-grain}

In our experiments, we measure the continuous position $x(t)$ in a double-well symmetric potential $U(x,t)$.  Because the energy barrier $E_{\rm b}$ of the double-well potential is much higher than $k_{\rm B}T$ for initial and final states, we can consider the system to be effectively a two-state system at those times, with the particle either in the left well (state $L$), defined by $x<0$, or the right well (state $R$), defined by $x>0$.  In this section, we derive the second law for such initial/final two-state systems, relating it explicitly to the underlying continuum description.

To accomplish this, we define the notion of local equilibrium in the potential $U(x,t)$, where, in the discussion below, $t$ is either the initial time $0$ or the final time $\tau$.  That is  the system is in state $L$ (left) with probability $p(t)$ and state $R$ (right) with probability $1-p(t)$.  But, constrained to be within one well or the other, the system is in thermal equilibrium.  

We can thus define a conditional equilibrium free energy $F_{\rm leq}(t)$, which is the free energy of the system given that it is in the left well \cite{junier09,sagawa14}.  In analogy with the usual definition of the equilibrium free energy, we have,
\begin{equation}
	F_{\rm leq}(t) = -k_{\rm B}T \, \ln Z_{\rm leq}(t) \,,
\label{eq:freeEnergyLocal}
\end{equation}
where the conditional partition function $Z_{\rm leq}(t)$ is given by integrating $\exp[-U(x,t)/k_{\rm B}T]$ over the interval $(0,\infty)$.  $F_{\rm leq}(t)$ is also known as the ``conformational'' free energy \cite{roldan14}.  Because of the assumed symmetry of the initial/final potential, $F_{\rm leq}(t)$  is the same if evaluated over the other state, $R$.  Otherwise, one would define local quantities for each state.  Notice that we can invert Eq.~\ref{eq:freeEnergyLocal} to write $Z_{\rm leq}(t) = \exp [-F_{\rm leq} (t)/ (k_{\rm B}T)]$.

We can then define a local-equilibrium density function,
\begin{align}
	\rho_{\rm leq}(x,t) &= \exp \left[ (F_{\rm leq}(t)-U(x,t)) / (k_{\rm B}T) \right] \nonumber \\ 
		&\qquad \times \left[ p(t) \, \theta(-x) + [1-p(t)] \, \theta(x) \right] \,,
\label{eq:rhoLocalEq}
\end{align}
where $\theta(x)$ is the Heaviside step function, 0 for $x<0$ and 1 for $x \ge 0$.  The physical meaning of $\rho_{\rm leq}(x,t)$ is that the particle is in local equilibrium in the left well of the potential $U(x,t)$  with probability $p(t)$ and in local equilibrium in the right well with probability $1-p(t)$.  

 Notice, too, that $\rho_{\rm leq}(x,t)$ is typically not the global equilibrium Boltzmann-Gibbs distribution associated with the potential $U(x,t)$, which would have $p(t)= \tfrac{1}{2}$.

 We next decompose the nonequilibrium density $\rho(x,t)$, using the law of total probability, into left and right components:
\begin{equation}
	\rho(x,t) =  p(t) \, \rho (x,t | x<0) + [1-p(t)] \, \rho (x,t | x>0) \,.
\label{eq:rhoLocalNeq}
\end{equation}
In contrast to the form given in Eq.~\ref{eq:rhoLocalEq}, the nonequilibrium $\rho$ makes no hypotheses as to the form of the conditional densities.  However, the function $p$ is chosen to be the same in both densities.  Because Eq.~\ref{eq:rhoLocalNeq} simply applies the definition of conditional probabilities, it is always possible to write the nonequilibrium density in this way.

Interpreting the entropy $S$ as the Gibbs-Shannon entropy associated to the nonequilibrium density, Eq.~\ref{eq:GibbsShannon}, the nonequilibrium free energy $F_{\rm neq}$ can be expressed in terms of the local equilibrium as
\begin{align}
	F_{\rm neq}(t) = F_{\rm leq}(t) &- k_{\rm B}T \, (\ln2) \, H[p(t)] \nonumber \\
		&+ k_{\rm B}T \, D_{\rm KL}\left[ \rho(x,t) \, || \, \rho_{\rm leq}(x,t) \right] \,,
\label{eq:Fneq}
\end{align}
where $H [p(t)]$ is the discrete binary Shannon entropy (in bits),
 \begin{equation}
	H(p) = -p \log_2 p - (1-p) \log_2 (1-p) \,,
\label{eq:ShannonInfo}
\end{equation}
and where the relative entropy (Kullback-Leibler divergence) is \cite{cover06}
\begin{equation}
	D_{\rm KL} [ p(x) \, || \, q(x) ] \equiv \int_{-\infty}^\infty {\rm d}x \,
		p(x) \, \ln \left( \frac{p(x)}{q(x)} \right) \,,
\label{eq:DKL}
\end{equation}
for probability density functions $p(x)$ and $q(x)$.  Equation~\ref{eq:Fneq} can easily be generalized to an asymmetric multi-well potential; particular cases are proved in \cite{shizume95,sagawa09,sagawa14}.  Note that for $0 < p < 1$, the Shannon entropy $H(p)$ ranges between 0 and 1 bit, and the relative entropy measures the distinguishability of two probability distributions and satisfies $D_{\rm KL} [ p(x) \, || \, q(x) ] \ge 0$, equaling zero only when $p(x) = q(x)$.  See the Supplement for a derivation of Eq.~\ref{eq:Fneq}.  The second law with discrete entropy is then found by combining Eqs.~\ref{eq:secondLaw3} and \ref{eq:Fneq}.  Note that the relative-entropy term quantifies the effect of the departure from local equilibrium in the second law, an issue that has been studied from a different point of view in Ref.~\cite{chiuchiu15b}.

\section*{Protocols for measuring the function $H(p)$ }

The main idea is that, for slow, thermodynamically reversible protocols, the inequality in Eq.~\ref{eq:secondLaw3} becomes an equality, giving with Eq.~\ref{eq:Fneq}  a way to obtain the function $H(p)$ experimentally.  To isolate the discrete entropy, we consider first a cyclic protocol that starts and ends with the system having the same symmetric double-well potential $U(x)$.  This eliminates the free-energy difference $\Delta F_{\rm leq}$. Moreover, we choose the initial density to always be in local equilibrium, and we choose protocol times $\tau$ that are large enough that the final protocol is in local equilibrium, too, in the potential $U(x)$.   (Of course, here and elsewhere in this paper, we always assume that the protocol time $\tau$ is shorter than the time to globally equilibrate via spontaneous hops over the barrier; that time scale is effectively infinite.) The relative entropy term in Eq.~\ref{eq:Fneq} then vanishes at both $t=0$ and $t=\tau$.  Finally, under these conditions, the change in non-equilibrium free energy is simply, from Eq.~\ref{eq:Fneq}, 
\begin{equation}
	\Delta F_{\rm neq} = - k_{\rm B}T \, (\ln2) \Delta H \,,
\label{eq:DFneq}
\end{equation}
This is the principle proposed by Landauer in 1961 \cite{landauer61} and studied extensively since \cite{bennett82,landauer91,shizume95,piechocinska00,plenio01,leff03,kawai07,sagawa09,dillenschneider09,maruyama09,esposito11,aurell12,jaksic14,sagawa14,chiuchiu15b,lutz15,parrondo15}, with recent experimental confirmation \cite{berut12,jun14,hong16,martini16,peterson16}.  Thus, by measuring the minimal average work to carry out protocols that alter the information content of a two-state system, we can test whether the Shannon entropy has physical relevance:  Does it apply to thermodynamic descriptions such as Eq.~\ref{eq:secondLaw3}? 
 
More precisely, we explore experimentally the two protocols illustrated in Fig. 1:

\begin{itemize}
\item \textit{Protocol 1}:  We erase completely a fraction of a bit of information.  The initial state of the system is a local equilibrium, with a probability $p_0$ for a particle to be in the left well.  The state encodes an information content $H_0 = H(p_0)$.  At the end of the protocol, at time $\tau$, the particle is again in local equilibrium but now always in the right well, implying that $H_\tau = 0$.  Thus, $\Delta H = -H_0$ and $\Delta F_{\rm neq} = k_{\rm B}T \, (\ln2) \, H_0$.

\item \textit{Protocol 2}:  We start with one bit of information and erase a fraction of it.  The initial state of the system is local equilibrium with $p_0 = \tfrac{1}{2}$, which corresponds to one bit of information.  The final state, after time $\tau$, is in local equilibrium with probability $p_\tau$ to be in the left well, corresponding to $H_\tau$ between zero and one bit.  Thus, $\Delta H = H_\tau - 1$  and $\Delta F_{\rm neq} = k_{\rm B}T \, (\ln 2) \, [1 - H_\tau]$. 

This protocol resembles that used in \cite{berut12,berut13,berut15}.  However, in those studies, partial erasure was used because the barrier could not be made high enough to ensure full erasure, and correction factors were applied to infer the work required for full erasure of a bit.  Here, we will use, in a controlled way, the partial work as a means to estimate the Shannon entropy function, $H(p)$.
\end{itemize}

\begin{figure}[tbh]
	\begin{center}
		\includegraphics[width=3.5in]{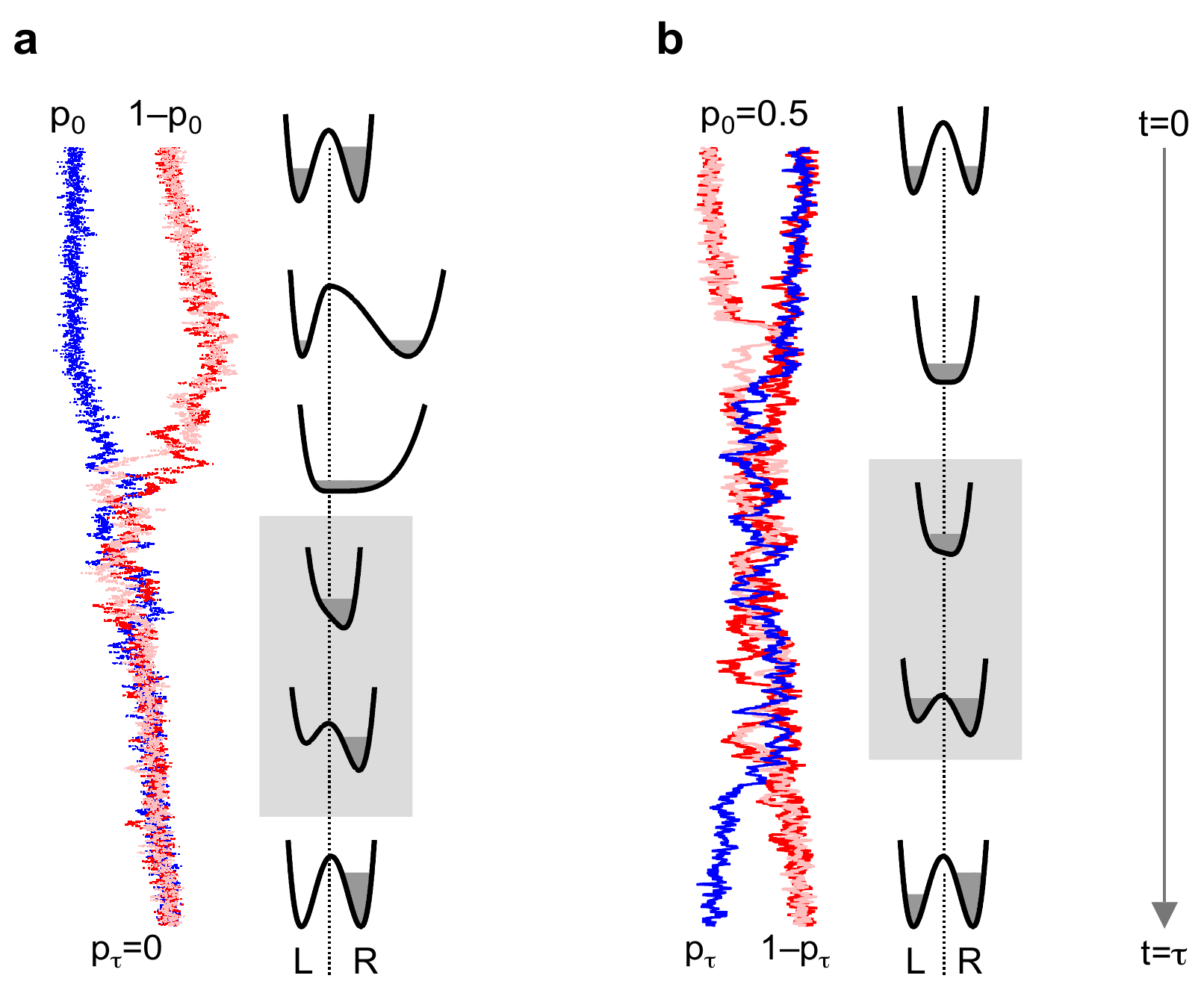}
	\end{center}
	\caption{Protocols of duration $\tau$ for erasing a fraction of a bit, accompanied by sample trajectories.  (A) Protocol 1: full erasure of a fractional bit. The potential is stretched to bring the two states to global equilibrium before mixing.  Full erasure is achieved using a strong tilt (gray shading).  One trajectory (blue) starts in the left well; two (red, pink) start in the right.  All end in the right well.  (B) Protocol 2: fractional erasure of a full bit.  Initial equilibrium state is mixed directly.  Weak tilt (gray shading) controls the final probability.  A quarter of the trajectories end in the left well.}
\label{fig:protocols}
\end{figure}

Figure~\ref{fig:protocols}a shows Protocol 1.  Naively, one might lower the barrier as a first step; however, such a protocol leads experimentally (and analytically) to an asymptotic work of $k_{\rm B}T \, \ln 2$ for all initial probabilities $p_0$ (see supplement).  But first stretching the potential to bring the system to global equilibrium before lowering the barrier allows it to reach the reversible bound, $k_{\rm B}T \, (\ln 2) \, H(p_0)$.  We thus stretch, lower the barrier, compress, strongly tilt, raise the barrier, and finally untilt to return the potential to its initial shape.  For a strong tilt, all observed trajectories end in the right well.

\section*{Results}

\begin{figure}[h]
	\begin{center}
		\includegraphics[width=3.5in]{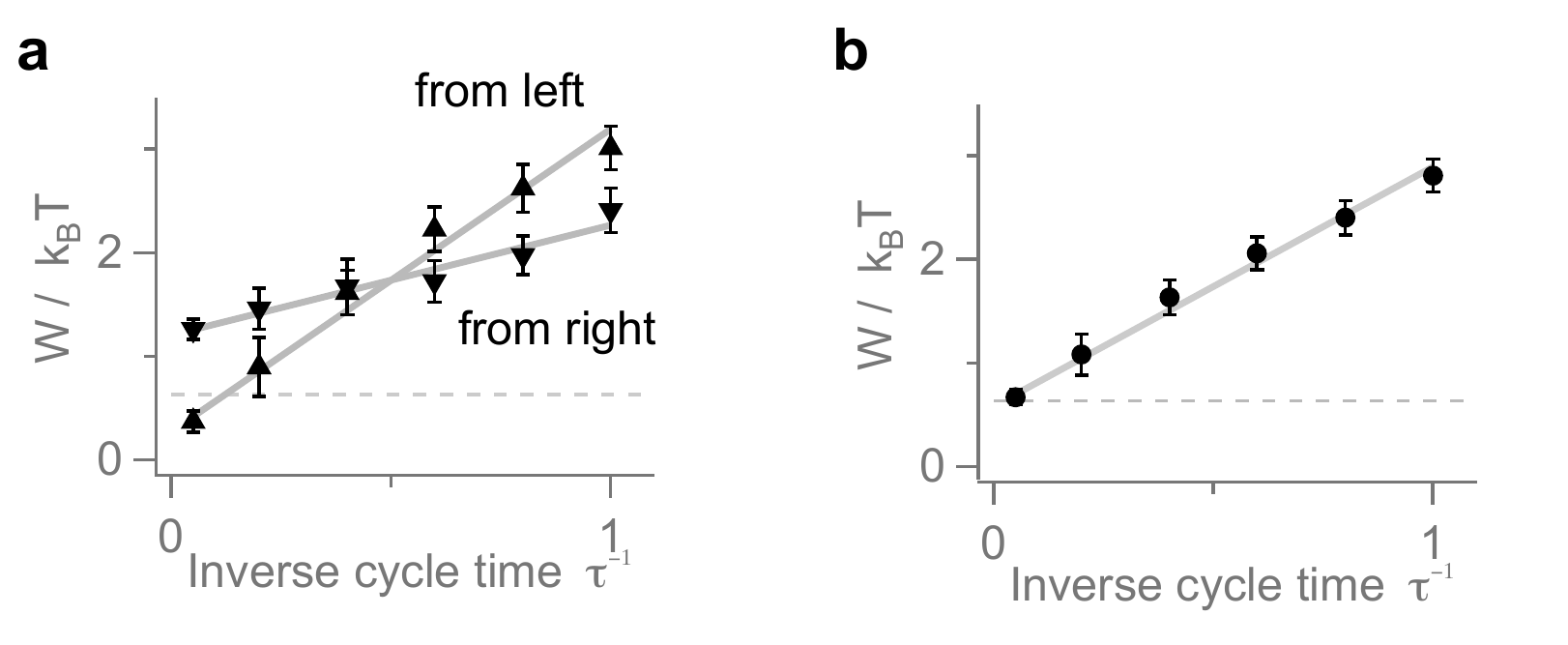}
	\end{center}
	\caption{Work to erase a fraction of a bit (Protocol 1).  (A) Conditional work measurements for particles starting in the left and right wells.  (B) Unconditioned work required to erase a fraction of a bit for $p_0=\frac{1}{3}$ at finite times $\tau$.  Extrapolating the fit gives $W_{\infty} / k_{\rm B}T = 0.58 \pm 0.07$, with $\chi^2 = 1.4$ for 4 degrees of freedom.  The dashed horizontal lines denote ($\ln 2$) times the change in information in bits: ($\ln 2) \, \Delta H = (\ln 2) \, H(\frac{1}{3}) \approx 0.64$, as calculated from Eq.~\ref{eq:ShannonInfo}.}
\label{fig:ProtocolWork1}
\end{figure}

\begin{figure}[tbh]
	\begin{center}
	\includegraphics[width=3.5in]{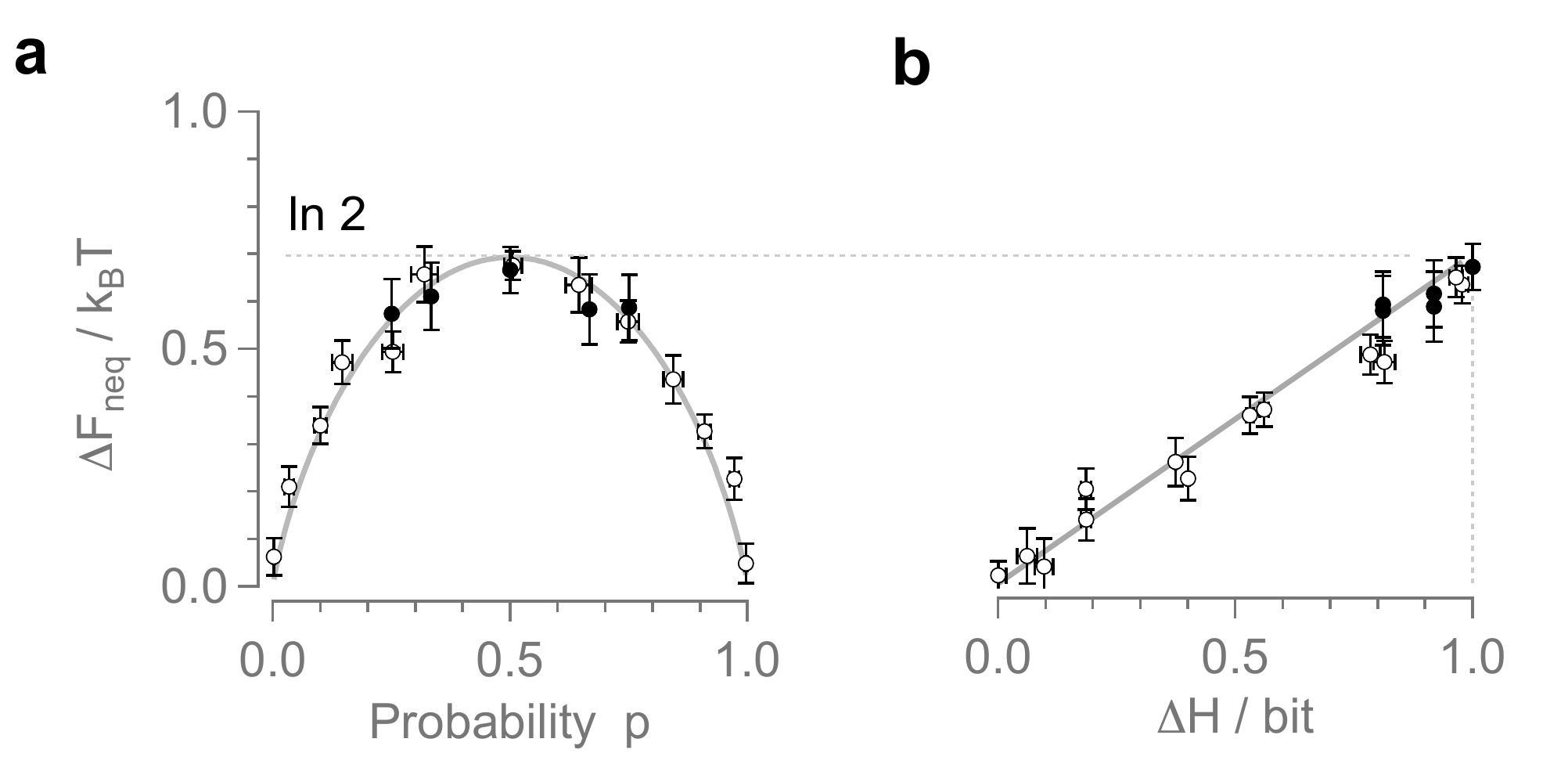}
	\end{center}
\caption{Change in non-equilibrium free energy due to a partial memory erasure.  Filled markers are measured using Protocol 1 by extrapolation, hollow markers using Protocol 2 at fixed cycle time $\tau=2$.    (A) Plot vs. probability, $p_0$ and $p_\tau$, respectively, in the two protocols.  Solid gray line is a plot of $H(p)$, with no fit parameters.  (B) Plot vs. change in Shannon entropy, in the limit of slow protocols.  Solid line---not a fit---shows the predicted slope of ln 2 $\approx 0.69$ per bit, from Eq.~\ref{eq:secondLaw3}.}
\label{fig:shannon2} 
\end{figure}

For each cycle time $\tau$ and each initial state, we find the average work.  Figure~\ref{fig:ProtocolWork1}a shows the average conditional  work for particles starting in the left  and right wells.  Figure \ref{fig:ProtocolWork1}b shows the combined average work.  Work in the slow limit is estimated by extrapolating to long times.  In this limit, the protocol is fully reversible, and the nonequilibrium free-energy change equals the work done by the potential, $\Delta F_{\rm neq}=W_{\infty}$.  We plot the scaled change in nonequilibrium free energy $\Delta F_{\rm neq} / k_{\rm B}T$ as a function of $p_0$ in Fig.~\ref{fig:shannon2}a. 

Our measurements show that it takes less than $k_{\rm B}T\ln2$ of work to erase less than one bit of information.  Although the results from Protocol 1 are consistent with the expected shape of the Shannon entropy function, $(\ln 2) \, H(p_0)$, they test only a narrow range of $p_0$, since large stretching factors $\eta$ imply long time scales ($\sim \eta^2$ because of diffusion).

To explore a wider range of information erasure, we therefore developed a second protocol that tilts rather than stretches the potential to create an energy difference between two local minima.  Tilting a potential does not increase its spatial extent and allows us to explore the full change of information from 0 to 1 bit.  However, there are problems that preclude extrapolating small-tilt protocols to long times (see supplement).

We thus designed a protocol that operates at a \textit{fixed}, large cycle time $\tau$.  At fixed $\tau$, the mean work $W(\tau)$ needed to change the information from $H_0$ to $H_\tau$  is always strictly greater than the change in free energy $W >\Delta F_{\rm neq}$ (Fig.~\ref{fig:ProtocolWork1}b).  To isolate the lower bound of the work, we run the protocol in the forward and then the backward direction.  When the protocol is executed slowly enough that conditional work distributions are Gaussian, we find (see supplement)
\begin{equation}
	\tfrac{1}{2} \left( W_{\rm F} - W_{\rm B} \right) = \Delta F_{\rm neq} 
		= k_{\rm B}T \, (\ln 2) \, [1 - H_\tau] \,,
\label{eq:avgWorkExtraction0}
\end{equation}
where $1 - H_\tau$ is minus the change in Shannon entropy and $W_{\rm F}$ ($W_{\rm B}$) the average work for the forward (backward) part of the protocol.  Similar formulas have been used to estimate \textit{equilibrium} free energy differences \cite{collin05,kim12}.   Here, we estimate the \textit{nonequilibrium} free energy difference using Eq.~\ref{eq:avgWorkExtraction0}.

Figure~\ref{fig:shannon2}a shows the results of Protocol 2 (hollow markers), plotted as $\ln 2 - \Delta F_{\rm neq} / k_{\rm B}T$ so that the data from Protocols 1 and 2 may be compared directly.  The plot agrees---without fit---with the Gibbs-Shannon form, $(\ln 2) \, H(p)$, over the full range $p \in [0,1]$.  Figure \ref{fig:shannon2}b then shows that this change in nonequilibrium free energy is linear in the Shannon entropy change.  

\section*{Discussion}

Although our focus in this paper is on testing the Gibbs-Shannon entropy for discrete states, Eq.~\ref{eq:ShannonInfo}, we measure a continuous position and can test explicitly aspects of the continuum version of the entropy, Eq.~\ref{eq:GibbsShannon}.  For example, we argue in the supplement that our data are consistent with a conditioned version of Crooks' relation.  Further, we discuss how the measurements presented here also confirm the identification between the total entropy production $S_{\rm tot}$ and the relative entropy between the forward and backward path measures.  

Beyond their role in justifying the underlying assumptions made in the field of stochastic thermodynamics, our results may aid continuing efforts to understand the role of information in nonequilibrium biological systems, where cells actively sense their environment and respond.  For example, we saw that a naive version of Protocol 1 was intrinsically irreversible and therefore unable to reach the ultimate thermodynamic bounds based on starting and ending states.  In  recently published work \cite{ouldridge17}, Ouldridge et al. argue that realistic biochemical networks similarly cannot reach these fundamental bounds.   In that case, the authors trace the extra dissipation to a failure to exploit all correlations generated between the measuring device and the physical system (receptors and readouts).  It will be interesting to study systematically the various classes of explanations for dissipation beyond the minimum levels reached here in a more-idealized kind of experiment.

\section*{Conclusion}

Two different protocols that each measure the minimal average work required to erase a fraction of a bit of information both confirm that the nonequilibrium system entropy of a colloidal particle in a controllable potential has a functional form consistent with that proposed long ago by Gibbs and Shannon.

\vspace{-10pt}

\subsection*{Experimental Setup}

A feedback (or Anti-Brownian ELectrokinetic, or ABEL) trap is a technique for trapping and manipulating small particles in solution \cite{cohen05b}.  The basic idea is to replace a trapping potential with a feedback loop:  in one cycle, one measures the position of a particle and then applies a force (created by an electric field) that pushes it back to the desired trapping point.  By the next cycle, thermal fluctuations have pushed the particle in a different direction, and a new restoring force is computed.  Feedback traps can also be used to place particles in a virtual potential, where the motion imitates a desired potential \cite{cohen05d,jun12,gavrilov13,jun14}.  

In the protocols described below, we take advantage of the nearly complete freedom to specify arbitrarily the shape of a virtual potential.  Thus, we can selectively lower the barrier, while keeping the outer part of the potential fixed.  Or, we can selectively stretch one well by a factor $\eta$ while  the other well is unchanged.  Such manipulations are not possible in erasure experiments based on optical tweezers \cite{berut12,berut13,berut15}, which limits the possible protocols in such cases.

The challenge with using feedback traps to measure work values to an accuracy $<~0.1~k_{\rm B}T$ is to calibrate forces accurately and to account for slow drifts in quantities such as the particle's response to an applied voltage.  In earlier work, we developed a recursive, real-time calibration technique \cite{gavrilov14} that allows us to measure accurately the stochastic work done by a changing potential on a particle.  Using an improved setup with higher feedback loop rates \cite{gavrilov15}, we explored erasure in asymmetric memories \cite{gavrilov16b}, tested subtle forms of reversibility \cite{gavrilov16a}, and compared different estimators of heat transfer \cite{gavrilov17}. 

The experimental setup for our feedback trap has three major segments: the imaging system, the trapping chamber, and the control software.  The imaging system consists of an inverted, home-built, dark-field, front-illumination microscope with a 60x Olympus NA=0.95 air objective \cite{weigel14,gavrilov15}.  A silica bead of diameter 1.5~$\mu$m is illuminated by a 660-nm LED source.  A small disk placed behind the objective blocks the direct LED light but allows scattered light to reach a camera.  The camera (Andor iXon DV-885) takes a $50 \times 20$ pixel image every $\Delta t$~=~5~ms, with an exposure $t_c$~=~0.5~ms.  The trapping chamber is cylindrical, $\approx10$~mm in diameter and $\approx5$~mm in height, and is glued on top of a glass coverslip.  We load silica beads diluted in deionized water.  The beads sink to the bottom of the chamber (top of the coverslip) under gravity, which confines them in the vertical ($z$)  direction.  Two pairs of electrodes near the bottom of the chamber create an electric field ($\sim$ 10 V/cm) whose value is updated every time step to move a bead in the horizontal ($xy$) plane \cite{gavrilov16a,gavrilov16b,gavrilov17}.   The control software analyzes images in real time using a centroid algorithm \cite{berglund08}.  It calculates forces based on the measured position and value of the gradient of the virtual potential.  Simultaneously, deviations between the expected and measured positions are used to calibrate the feedback trap, using a recursive maximum likelihood algorithm for a continuous linear fit between the applied voltages and observed displacements \cite{gavrilov14}.  The particle's electric-field mobility is estimated from the slope.  Drifts are assessed via the intercept, and particle diffusion is estimated from the fit residuals.  A running-average algorithm keeps only the most recent measurements and helps track parameter changes during experiments that can last several days.

Finally, in the supplement, we justify in more detail our model of the dynamics as one dimensional and overdamped \cite{happel83}.

\subsection*{Data Analysis}  
Feedback traps allow one to impose an arbitrary virtual potential of almost any form.  We choose a static harmonic potential in the $y$ direction and a double-well potential in the $x$ direction:
\begin{equation}
	U(x,t) = 4 E_{\rm b} \left[ -\tfrac{1}{2} g(t) \, \tilde{x}^2 + \tfrac{1}{4} \tilde{x}^4 
		- A f(t) \, \tilde{x} \right] \,,
\label{eq:DWpotential}
\end{equation}
where the scaled coordinate $\tilde{x}(x,t)$ is selectively stretched for positive or negative $x$, as desired.  More precisely, $\tilde{x}(x<0,t) = - \tilde{\eta}(t)\tilde{x}(x\geq0,t)$ where $\tilde{\eta}(t)$ is a time-dependent stretching factor (see Fig.~\ref{fig:protocols}a).  Note that the stretching amplitude $\eta$ scales the stretching factor $\tilde{\eta}(t)$.  In all cases, $\tilde{\eta}(0) = \tilde{\eta}(\tau)=1$, so that we start and end with a symmetric potential. In Eq.~\ref{eq:DWpotential}, $E_{\rm b}$ is the energy barrier height,  $A$ the tilt amplitude.  The functions $g(t)$ and $f(t)$ can take values between $0$ and $1$ and control the barrier height and tilt.  Together with stretching $\tilde{\eta}(t)$, they allow us to implement Protocols 1 and 2, as described below and in the Supplement.

Each experiment uses several beads, whose properties must each be measured using the recursive algorithm given above.  Via dimensionless scaling, we can combine data measured on beads, which, although nominally identical, differ slightly in radius and charge.  The measured diffusion constant near the surface is typically $\approx$ 0.23 $\mu$m$^2/$s.  Based on the requirement that feedback update time be much smaller than the local relaxation time within a well, we set the distance between two local minima of the double-well potential.  A typical value is $2x_0$ = 1.54 $\mu$m.  The dimensionless time $\tau=1$ then corresponds to a physical time $t_{\rm sec}\approx10$~s.

The work to manipulate a potential in one cycle of duration $\tau$ is estimated by discretizing  Sekimoto's formula \cite{sekimoto97,sekimoto10} for the stochastic work, 
\begin{equation}
	w_{\tau} = \int_0^{\tau} {\rm d}t \, \left. \frac{\partial U(x,t)}{\partial t} \right|_{x=x(t)} \,.
\label{eq:sekimotoWork}
\end{equation}

\subsection*{Experimental Protocols} 
We used two different erasure protocols.  In both, we prepare the initial state by placing a particle in a given well using a strong harmonic trap for $0.5$~s.  We then abruptly switch to a static double-well potential to let a particle equilibrate locally for $1$~s, before the cycle starts.  Below, we describe qualitatively each protocol.  (See Supplement for the explicit potentials, $U(x,t)$.)

\subsubsection*{Protocol 1} The initial state is in the left well with probability $p_0$ and has system entropy $H_0 = H(p_0)$.  We erase to a state with $p_\tau =0$ (always in the right well) and $H_\tau =0$.  We define the initial state of the memory by placing a particle in a particular well.   The high energy barrier of $E_{\rm b} = 13~k_{\rm B}T$ prevents the two states from mixing on the time scales of the experiment.

We measure the mean work for full erasure from this initial state via conditional work values.  That is, we measure the average value of work $W_{\rm L}$ to erase conditioned on starting in the left well and similarly for the right well, $W_{\rm R}$.  For $N_{\rm L}$ individual measurements $w^i_{\rm L}$, we estimate the mean via the average, $W_{\rm L} \approx \overline{W}_{\rm L} = \frac{1}{N_{\rm L}}\sum_i w_{\rm L}^i$.  Similarly, $W_{\rm R} \approx \overline{W}_{\rm R} = \frac{1}{N_{\rm R}}\sum_i w_{\rm R}^i$.  The unconditional work at time $\tau$ is estimated from the law of total probability as $W_\tau = p_0 W_L + (1-p_0) W_R$.  The work in the slow limit is obtained by extrapolating using the asymptotic form $W_{\tau} \sim W_{\infty}+a\tau^{-1}$ and fitting a line against $\tau^{-1}$ \cite{sekimoto97a,aurell12}.

We need to start by stretching the potential by a factor $\eta = 1/p_0 - 1$, to equalize the probability densities in the left and right states and bring them to global equilibrium.  Otherwise, lowering the barrier would be an irreversible step that adds dissipation that does not vanish, even in the slow limit \cite{gavrilov16a}.

For $p_0 \geq 0.5$, the left well is stretched, while, for $p_0 < 0.5$, the right well is stretched.  (At $\eta=1$, the wells have their minimum width, a width set by requiring that gradients be small enough that the discrete approximation to a continuous potential is accurate \cite{jun12}.  We thus stretch one or the other well, depending on $p_0$.)  Note that, as a consequence of the stretching, the values of $W_{\rm L}$ and $W_{\rm R}$ depend on $p_0$.  After stretching, we lower the barrier and mix the states, then strongly tilt towards the right.  Finally, we increase the barrier and untilt the potential.  This protocol is repeated for several different cycle times $\tau$, where, for each $\tau$, we recorded multiple trajectories over a thirty-minute period.  The uncertainty in the estimate of average work values depends only on the total time of data collection, not on the cycle time $\tau$ directly \cite{gavrilov17}.

\subsubsection*{Protocol 2} The initial state has one bit of information, which is erased partially.  The initial state at time $t=0$ is in global equilibrium, with $p_0 = 0.5$ and $H_0=1$ bit, and ends with $H_\tau$, which we control in the range from 0 to 1 bit.  The slightly lower energy barrier $E_{\rm b} = 10~k_{\rm B}T$ reduces the distance between wells, which must be large enough that the virtual potential lead to dynamics that are indistinguishable from those of the corresponding physical potential \cite{jun12}.  Because the fixed cycle time is short ($\approx 30$ s), the probability of a spontaneous hop over the barrier is negligible.

Protocol 2 operates at the fixed cycle time $\tau = 2$.  In four steps, we lower the barrier and mix states, apply a weak tilt with an amplitude $A$, raise the barrier, and untilt.  The entire protocol is then repeated in reverse.  For each tilt $A$, we acquire data for about 12 hours.  We measure the stochastic work from each trajectory and the probability to end in the left well $p_\tau$ after the forward protocol.  As a control,  we estimate the probability to end in the left well after reverse protocol, which is consistent with the expected value of 0.5 for a reversible protocol.  (See supplemental material for data.)

Ensemble averages for Protocol 2 are estimated from the arithmetic mean of $N$ work measurements in the forward and reverse protocols:  $W_{\rm F} \approx \overline{W}_{\rm F} = \frac{1}{N}\sum_i W_{\rm F}^i$ and $W_{\rm B} \approx \overline{W}_{\rm B} = \frac{1}{N}\sum_i W_{\rm B}^i$.  By recording the work done for forward and backwards protocols at a fixed cycle time $\tau$, we have a simple, accurate way to estimate the change in nonequilibrium free energy (see section 4 of supplement).  Error bars on work measurements in all cases represent the standard error of mean, calculated as $\sigma_W/\sqrt{N}$, with $\sigma_W$ the standard deviation of the $N$ individual measurements.

\section{Acknowledgment}
\begin{acknowledgments}
We thank J.~Vollmer, D. Sivak, \'E.~Rold\'an, H.~Touchette, and M.~Esposito for suggestions.  This work was supported by a Discovery Grant from NSERC (Canada).  R.C.~was supported by the French Ministry of Education through grant ANR-15-CE40-0020-01.
\end{acknowledgments}

\clearpage
\setboolean{@twoside}{false}
\includepdf[pages=1]{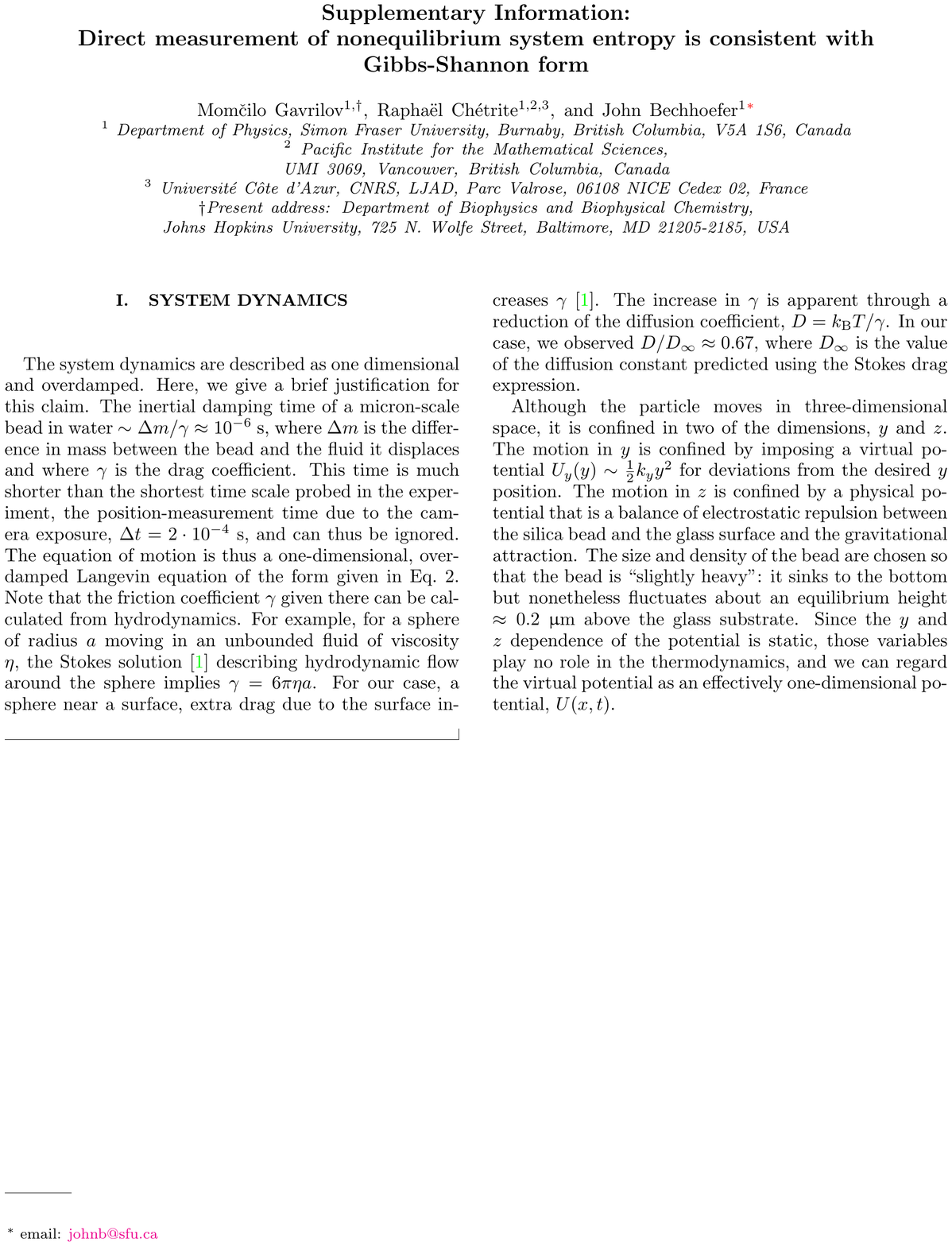}
\clearpage
\setboolean{@twoside}{false}
\includepdf[pages=2]{ShannonSupp.pdf}
\clearpage
\setboolean{@twoside}{false}
\includepdf[pages=3]{ShannonSupp.pdf}
\clearpage
\setboolean{@twoside}{false}
\includepdf[pages=4]{ShannonSupp.pdf}
\clearpage
\setboolean{@twoside}{false}
\includepdf[pages=5]{ShannonSupp.pdf}
\clearpage
\setboolean{@twoside}{false}
\includepdf[pages=6]{ShannonSupp.pdf}
\clearpage
\setboolean{@twoside}{false}
\includepdf[pages=7]{ShannonSupp.pdf}
\clearpage
\setboolean{@twoside}{false}
\includepdf[pages=8]{ShannonSupp.pdf}
\clearpage
\setboolean{@twoside}{false}
\includepdf[pages=9]{ShannonSupp.pdf}
\clearpage
\setboolean{@twoside}{false}


\section*{Supplementary material}

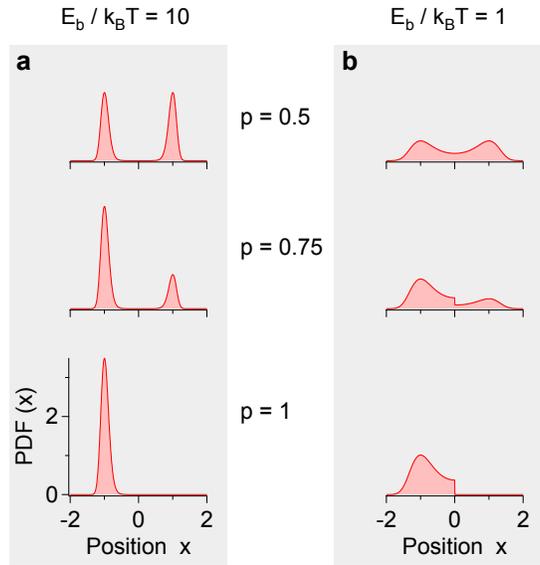

FIG. S1. Effect of dimensionless barrier height $E_{\rm b}/k_{\rm B}T$ on local equilibrium distribution, $\rho_{\rm leq}(x)$ for a symmetric, double-well potential. **a**, For a high barrier, there are effectively two separate distributions for all weights $p(t)$. **b**, For a low barrier, there is a significant jump discontinuity in density at $x = 0$ for $p(t)$ sufficiently different from 0.5.

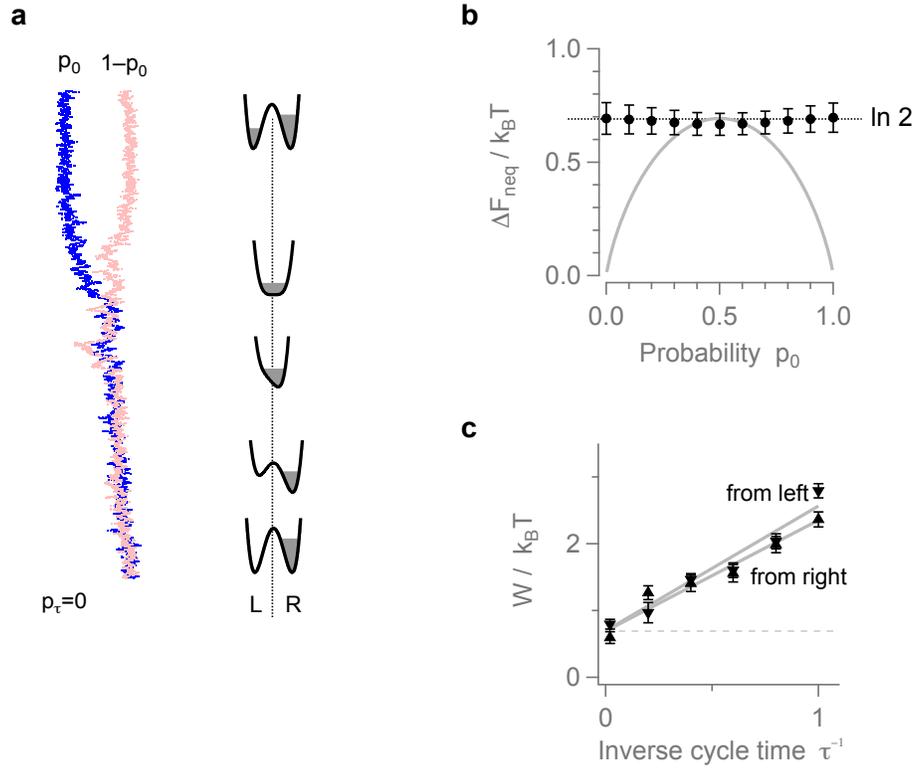

FIG. S2. Naive version of Protocol 1. **a**, Sample trajectories starting from left and right wells, along with potentials at time points within the protocol. **b**, For all initial probabilities $p_0$, the average change in nonequilibrium free energy to erase is $k_{\rm B}T \ln 2$ for slow protocols. **c**, The average conditional work to erase is asymptotically $k_{\rm B}T \ln 2$ for particles that start in the left well and also for particles that start in the right well.

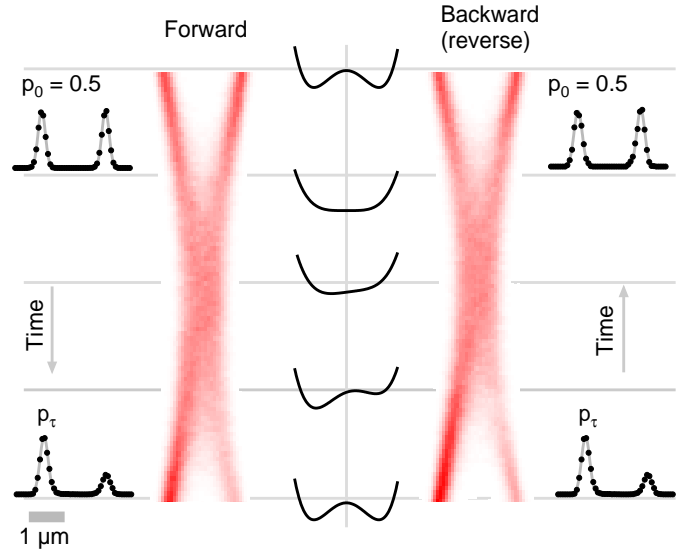

FIG. S3. Protocol 2: Path probability densities for partial erasure in the forward protocol (left) and its accompanying backward protocol (right). In the forward protocol, one bit of information is erased to the left well, with probability $p_\tau = 0.75 \pm 0.02$ for tilt amplitude $A = -0.03$. The duration of the protocol, $\tau = 2$, corresponds to a physical time of 20 s. The backward protocol, played forward in time, returns a particle to the initial state with probability 0.5.

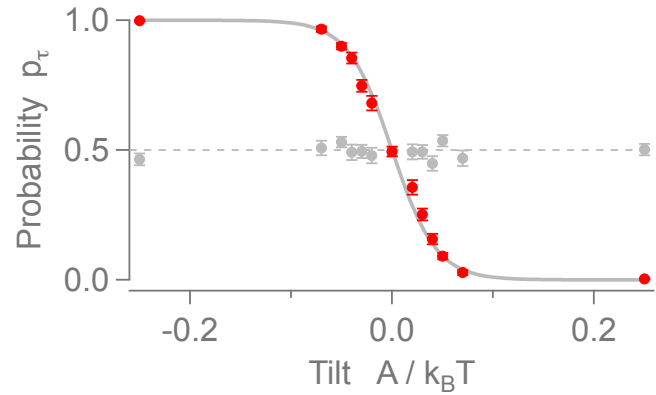

FIG. S4. Erasure probability recorded for different tilt amplitudes. Red markers show probability $p_\tau$ at the end of the partial erasure experiment, while gray markers show the probability of ending up in the left well for the time-reversed protocol. Solid gray line is empirical function relating the tilt amplitude $A$ to the probability $p_\tau$ of being in the $L$ state at time $\tau$, given by $p_\tau = f(A) = 0.5[1 - \tanh(23A)]$ for $\tau = 2$.

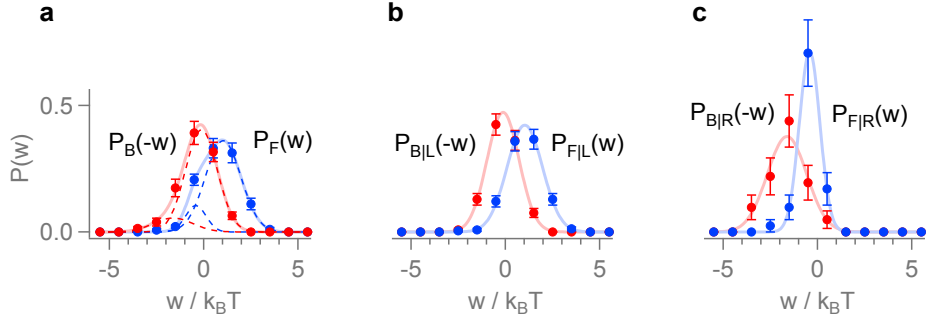

FIG. S5. Conditional work distributions are consistent with Gaussian. **a**, Estimated unconditioned work distributions $P_\mathrm{F}(w)$ and $P_\mathrm{B}(-w)$ for forward and reverse parts of protocols (red and blue markers). Solid lines represent the sum of the corresponding two Gaussian distributions in b and c. Dashed lines show contributions from weighted conditional Gaussian distributions. **b**, Conditional work distributions for the left state for forward $P_\mathrm{F|L}(w)$ and backward $P_\mathrm{B|L}(-w)$ protocols. Solid lines are fits to Gaussian distributions. **c**, Same, for right state. Tilt amplitude is set to $A/k_\mathrm{B}T = -0.04$, and we measure $p_\tau = 0.85 \pm 0.02$.

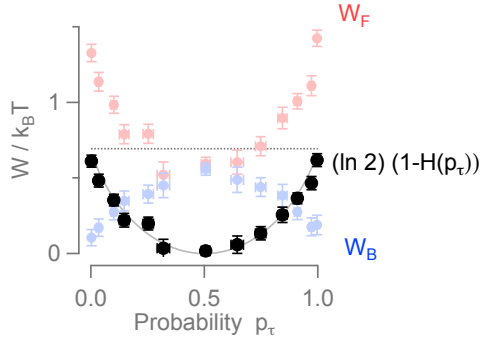

FIG. S6. Mean work in forward protocol $W_\mathrm{F}$ compared with reverse protocol $W_\mathrm{B}$. The desired Shannon-entropy term $H(p_\tau)$ is isolated from $W_\mathrm{F}$ and $W_\mathrm{B}$ using Eq. 14 of the main text, or Eq. S24 here.

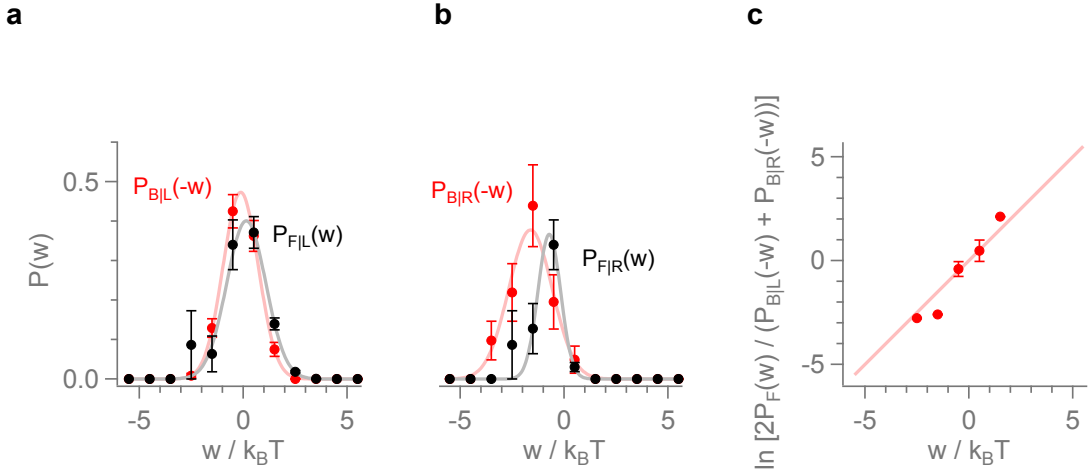

FIG. S7. Comparison between measured and calculated conditional work distributions for Protocol 2. **a**, Conditional work distribution with left-well conditioning. Red markers show measured values, black markers shows estimates using Eq. S12a. **b**, Conditional work distribution with right-well condtioning. Red markers show measured values, black markers shows estimates using Eq. S12b. **c**, Test of Eq. S26. Solid line has slope = 1.



\end{document}